\documentstyle[aps,pra,epsfig]{revtex}
\begin{document}
\title{The Markov approximation for the atomic output coupler}  
\author{M. W. Jack,$^{1}$ M. Naraschewski,$^{2}$ M. J. Collett,$^{1}$ and  
D. F. Walls$^{1}$}
\address{$^{1}$ Department of Physics, University of Auckland, Private Bag 92019, Auckland, New 
Zealand.\\
$^{2}$ Jefferson Laboratory, Department of Physics, Harvard University, Cambridge MA 02138\\
Institute for Theoretical Atomic and Molecular Physics,
Harvard-Smithsonian Center for Astrophysics, Cambridge MA 02138.}
\maketitle

\begin{center}
(PREPRINT: \today)
\end{center}
\begin{abstract}
The regions of validity of the Markov approximation for  
the coupling of atoms out of an atomic trap are determined. We consider radio-frequency output coupling in the presence of gravity and collisional  repulsion, and  Raman output coupling. The Markov approximation is crucial in most
theoretical descriptions of an atom laser that assume a continuous process
of output coupling from a trapped Bose-Einstein condensate. In this regime many
techniques proved to be useful for modeling the optical laser, such as master equations, can be used  to describe the dynamics of the damping of the condensate mode undergoing output coupling. 
\end{abstract}
\vspace{1cm}  
                           
\section{Introduction} \label{introduction}
The recent observation of Bose-Einstein condensation (BEC) in atomic
traps \cite{anderson,davis} has attracted widespread attention. One of
the most important perspectives of this experimental achievement is
the possibility of producing the matter wave analog of a laser, i.e.,
a high flux source of coherent atoms. In its simplest form, such an
atom laser can be built by adding a suitable output coupling mechanism \cite{mewes,raman}
to present Bose-Einstein condensation experiments. Several attempts
have already been made to develop a theoretical description of an
atom laser \cite{hope97,moy,savage,hope98,ballagh,naraschewski97,zhang,guzman,holland,martin,wiseman,steck,kneer} that combines elements of kinetic theory
and laser theory.  One of the problems that exists in applying optical
laser concepts to the Bose-Einstein output coupler situation is the
question of the validity of the Markov approximation \cite{hope97}.
The Markov approximation is an extremely powerful tool to describe the
coupling of a system of trapped particles to an
environment.  It allows one to think of the coupling
in the following terms: at any one time, a particle is either in the trapped
system or has been coupled out. Quantum mechanically, this means that the existence of any 
superposition of these two possibilities is neglected. Such an assumption is
valid if the superposition decays on a time scale much faster than
changes occur in the state of the trapped system. Classically, the Markov approximation implies  that an atom that has been coupled 
out will have no chance of being  brought back into the trap again. 
In the case of a noninteracting gas and ignoring gravity, atoms that are coupled out leave the spatial region of the trapped atoms  due to the 
relatively slow quantum spreading of their wave packet. As a consequence, the superposition decays slowly and it is likely that some atoms will be coupled back into the trap instead of leaving
the system irretrievably. Such a behavior leads to a strongly non-Markovian
dynamics of the condensate mode \cite{hope97,moy,savage,hope98}. 
However, in the presence
of accelerating potentials atoms may be removed from the region of coupling
at a much faster rate and thereby allow the use of the Markov approximation.

In general, there are two distinct operating regimes for an output
coupler. The
first regime is the strong-coupling regime, where portions of a
condensate are coupled out at such a rate that the output-coupled
atoms do not have time to propagate while the coupling is in
progress. This  regime was realized in the first 
experimental demonstration of an output
coupler \cite{mewes}, where a rf-pulse was used to couple out a large fraction
of the trapped atoms within a time
interval of the order of $\mu s$.  When the coupling time was long compared to the rate at which atoms can be coupled back into the trap, strongly non-Markovian behavior, such as Rabi oscillations, was observed. In this regime strong collisional interactions dominated
the dynamics of the output-coupled atoms. The Gross-Pitaevskii equation (GPE)
has been found useful in numerically modeling such a situation
\cite{ballagh,naraschewski97,zhang}. However, in this paper
we are interested
in the opposite limit of weak semi-continuous output coupling, where
the untrapped atom beam is of low enough
spatial density that we can neglect the effect of collisions on its behavior. Output coupling in this
regime has not yet been demonstrated, although it is considered
experimentally viable \cite{mewes}.

Previous theoretical treatments have implicitly made use of the Markov
approximation by assuming a Lindblad master equation (see
Ref.\,\cite{louisell}) for the condensate mode
\cite{guzman,holland,martin,wiseman}. A different, though essentially
equivalent, approach treats the loss of condensate atoms by adding a
damping term to the time-dependent Gross-Pitaevskii equation
\cite{steck,kneer}. In the present work we explore the regimes
where the Markov approximation is valid. In Sec. \ref{formalisim} we
present the Hamiltonian of the total system and introduce the concepts
of a memory function and a memory time. In Sec. \ref{rf} we apply the
general ideas to the specific case of the radio-frequency output
coupler. The results obtained are summarized in
Sec. \ref{rfsummary}. In Sec. \ref{ramansec} we also address the case of a Raman
output coupler.

\section{Mathematical Formalism}    \label{formalisim}                       
The Hamiltonian of the total system can be written as the 
sum of three parts
\begin{equation}
        H=H_{T}+H_{U}+H_{I},
        \label{total hamiltonian}
\end{equation}
where $H_{T}$ is the Hamiltonian of the trapped particles, $H_{U}$ is the 
Hamiltonian of the untrapped particles and $H_{I}$ the interaction between the 
two. The Hamiltonian of the trapped atoms is of the form 
\begin{eqnarray}
    H_{T} & = & \int d{\bf x} \,\,\hat{\psi}_{T}^{\dagger}({\bf 
        x})\left[V_{T}({\bf x})+\frac{{\bf p}^{2}}{2m}\right]\hat{\psi}_{T}({\bf x})+ \frac{U}{2}\int d{\bf x} \,\,\hat{\psi}_{T}^{\dagger}({\bf x}) \hat{\psi}^{\dagger}_{T}({\bf x})\hat{\psi}_{T}({\bf x}) \hat{\psi}_{T}({\bf x}),
\end{eqnarray}
where  $\hat{\psi}_{T}({\bf x})$  is the field operator for the trapped atoms and  $V_{T}({\bf x})$ is the trap potential. $m$ is the mass of the atoms and $U=4\pi \hbar^{2} a/m $ is the coupling constant for a local collision process, where $a$ is the scattering length of a trapped-trapped collision.
The untrapped atoms are assumed to be of low enough spatial density that we can ignore collisions between the untrapped atoms and can write
\begin{equation}
 H_{U} =      \int d{\bf x} \,\,\hat{\psi}_{U}^{\dagger}({\bf 
        x})\left[V^{\rm eff}_{U}({\bf x})+\frac{{\bf p}^{2}}{2m}\right]\hat{\psi}_{U}({\bf x}),
\end{equation}
where $\hat{\psi}_{U}({\bf x})$ is the  field operator of the untrapped atoms and
$V^{\rm eff}_{U}({\bf x})$ is the effective non-confining potential experienced by the untrapped atoms. The form of this effective potential will be given in a later section. Both fields satisfy Bose commutation relations, 
$[\hat{\psi}({\bf x}),\hat{\psi}^{\dagger}({\bf x'})]=\delta({\bf x-x}')$. We consider situations where the output coupling is linear and conserves 
particle number,
\begin{equation}
        H_{I}  =      i\hbar\sqrt{\gamma}\int
        d{\bf x}\,\,\left[g({\bf x},t)\hat{\psi}_{U}^{\dagger}({\bf 
        x})\hat{\psi}_{T}({\bf x})-g^{*}({\bf
        x},t)\hat{\psi}_{U}({\bf x})\hat{\psi}_{T}^{\dagger}({\bf x})\right],
        \label{interaction hamiltonian}
\end{equation}
where the  coupling constant, $g({\bf x},t)$, is normalized so that
$\int d{\bf x}|g({\bf x},t)|^{2}=1$, and we assume that the strength of the coupling, $\sqrt{\gamma}$, is time independent. The time dependence of $g({\bf x},t)$ is then simply an oscillatory phase, $g({\bf x},t)=g({\bf x})e^{-i\nu t}$.   The above interaction Hamiltonian can describe radio-frequency output coupling \cite{mewes}, where a radio wave induces a transition from a hyperfine level that is trapped in the magnetic trap to one that is untrapped or anti-trapped. It can also describe
 Raman output coupling \cite{raman} where two laser beams cause an atom in the trap to make a two-photon transition to an untrapped state. In this case the atom experiences a momentum kick equal to the difference in the momentum of the photons involved in the transition.

In this work we confine our interest to  a single 
energy mode of the trapped system. We will concentrate on the case when the trapped mode of interest is the condensate mode. However, the method is equally applicable to an excited mode of the trap. 
 The trap mode operator of interest is defined in terms of its spatial mode function, $u_{a}({\bf x})$, by
\begin{equation}
        a\equiv\int d{\bf x}\,\,u_{a}^{*}({\bf x})\hat{\psi}_{T}({\bf x}).
        \label{modea}
\end{equation}
An oscillation frequency, $\mu$, will be associated with this mode. If the mode is the condensate mode then $u_{a}({\bf x})$ is
 the solution to the time-independent GPE and $\mu$ is the chemical potential. The mode operator $a$ is coupled to the untrapped field throughout an effective coupling region given by
\begin{equation}
        \kappa({\bf x},t)=g({\bf x},t)u_{a}({\bf x}).\label{kappa}
\end{equation}

\subsection{Damping of the trapped mode}\label{damping}
By substituting the formal solution of the Heisenberg equation for the 
untrapped field into the Heisenberg equation for the trapped mode $a$,
we obtain the Langevin equation of motion \cite{qn}, 
\begin{equation}
        \frac{da(t)}{dt}=-\frac{i}{\hbar}[a(t),H_{T}]-\gamma\int^{t}_{-\infty}ds f_{\rm  m}(t-s)a(s)-\sqrt{\gamma}\xi(t).
        \label{langevineqn}
\end{equation}
The driving field, $\xi(t)$, is the contribution from the free propagation of the initial untrapped field, considered here to be in a vacuum state,
\begin{equation}
\xi(t)\equiv\int d{\bf x} \,\kappa^{*}({\bf x},t)\hat{\psi}^{0}_{U}({\bf x},t),
\end{equation} 
where the dynamics of the free untrapped field operator $\hat{\psi}^0_U$ are determined by the Hamiltonian $H_U$ alone.
The presence of the driving field is necessary to preserve the bosonic commutation relations of the mode operator $a$.
The damping term (the second term on the right-hand-side) represents a loss of particles from the trapped mode into the untrapped field and makes explicit the dependence of the trapped mode on its past behavior via the so called memory function, $f_{\rm m}(t-t')$. The correlation between the driving field and itself at an earlier time determines the memory function via the commutation relation, 
 \begin{equation}
        f_{\rm m}(t-t') \equiv  [\xi(t),\xi^{\dagger}(t')].
         \label{memory function}
 \end{equation}
This relationship between the driving field and the memory function 
is an example of the quantum fluctuation-dissipation relation and leads to a description of a damping process consistent with both quantum and statistical mechanics.  The damping process can then be interpreted as the coupling of discrete atoms out of the trap at random times.

In general, the Langevin equation (\ref{langevineqn}) will
contain a term representing a free oscillation and other terms due to collisions 
between the trapped modes. We are interested in output coupling and the collisional behavior of the trapped atoms is not explicitly
modeled in this work. Instead,  we define a rate $\Gamma$  to account for these other processes 
without considering them explicitly. This rate may be calculated in work concentrating on the trapped atoms, such as \cite{gardiner}.  

\subsection{A finite memory time}\label{memory times}

For a dissipative system we expect the system behavior at time $s$ as $s\rightarrow -\infty$ to become less and less important in determining the present behavior of the system. To make this more concrete we define a memory time, $T_{\rm m}$, as the  time after which we can neglect the effect of the previous behavior of the system on the evolution in the present.
In terms of the Langevin equation, a memory time exists for the system if, at some finite time in the past, $T_{\rm m}$, we can make the approximation
\begin{equation}
 \int^{t}_{-\infty} dsf_{\rm m}(t-s)a(s) \simeq \int^{t}_{t-T_{\rm m}} dsf_{\rm m}(t-s)a(s), 
\label{finitememorycondition}
\end{equation}
for all $t$.   If this condition is satisfied then we will call $T_{\rm m}$ the memory time of the system. Note that if a memory time cannot be defined for the system then the separation of the total system into a localized system interacting with an environment becomes inappropriate as there will be no time at which one can say that a particle has left the localized system and entered the environment.

In general, the memory time as given by Eq.\,(\ref{finitememorycondition}) depends on the nature of $a(s)$ and so no general statements can be made concerning this condition without detailed knowledge of the behavior of the trapped atoms. However, from the motion of the untrapped atoms one can determine cases  where condition Eq.\,(\ref{finitememorycondition}) can be satisfied without assuming too much about the behavior of the trapped system. 
This is clearer if we write the memory function in terms of the single particle Green's function as,
\begin{equation}
 f_{\rm m}(t-t')=\int d{\bf x}d{\bf x}'\: 
        \kappa^{*}({\bf x},t)\kappa({\bf x'},t') G({\bf x},t;{\bf x}',t'),\label{overlap}
\end{equation}
where 
\begin{eqnarray}
        G({\bf x},t;{\bf x}',t') & \equiv & [\hat{\psi}^{0}_{U}({\bf 
        x},t),\hat{\psi}_{U}^{0\dagger}({\bf x}',t')],\\
        & = & \langle \{0\}|\hat{\psi}^{0}_{U}({\bf x},t)\hat{\psi}_{U}^{0\dagger}({\bf 
        x}',t')|\{0\}\rangle,
        \label{greensfunction}
 \end{eqnarray}
is the single particle Green's function for the free atoms for $t\geq t'$ and $\kappa({\bf x},t)$ is the effective interaction region given by Eq.\,(\ref{kappa}). The memory function can therefore be interpreted as the overlap between an atom, with an initial wave packet of the shape of the interaction region, $\kappa({\bf x'},t')$,  with a wave packet, $\kappa^{*}({\bf x},t)$ after it has propagated for a time $t-t'$.
There are two distinct ways in which this overlap could become smaller with increasing time.
Firstly, if an atom is leaving the interaction region due to an accelerating potential or quantum mechanical spreading of its wave packet then the overlap between the atoms wave packet and its original wave packet will decrease in time. The time at which we can neglect this overlap will determine the memory time. 
Secondly, if an atom is accelerating then it will gain kinetic energy which will cause the wave packet of the atom to oscillate. The overlap between this wave packet and a stationary one will then also oscillate. This oscillation will average to zero when integrated over time scales much longer than the period of oscillation. In our case, we are interested in integrating over time scales short compared to the coupling time scale, $\gamma^{-1}$. The memory time can be defined as the time  when the oscillation is much faster than $\gamma$, as the overlap will average to zero after this time.  Another, equivalent, way of thinking about this is that only a range of frequencies will be close to resonance with the coupling. Particles with energies far from resonance will not be coupled back into the trap.

Following from these considerations we can make a more practical definition of a  memory time in terms of the motion of the untrapped atoms alone by
\begin{equation}
\left|\int^{t}_{-\infty} dsf_{\rm m}(t-s)\right|  \gg 
\left|\int^{t-T_{\rm m}}_{-\infty} dsf_{\rm m}(t-s)\right|, 
\label{abs}
\end{equation}
independent of $a(t)$. Note that we have made the replacement $f_{\rm m}(t-t')\rightarrow f_{\rm m}(t-t')e^{-i\mu t'}$ to take account of the oscillation of the mode $a(t)$, as this may cancel the oscillation of the memory function itself. This is the only aspect of the behavior of $a(t)$ that we will consider in determining a memory time.

In summary, there is a certain region of phase space in the untrapped field where particles can be coupled back into the trap. The time taken for a particle to leave this region of phase space determines the memory time.   The memory-time can then be interpreted as the time interval after which we can safely assume that a particle has irretrievable left the trap.

If a change in $a(t)$ occurs during the memory time then it is necessary to consider the coupling out of an additional particle before the first particle has either been coupled back into the trap or left irretrievably; this is the non-Markovian regime. Strictly speaking we should distinguish between the free evolution of $a(t)$ and its evolution due to the output coupling. Strong coupling, $\gamma T_{\rm m}> 1$, leads to the second-order effects mentioned above and this is a serious breakdown of the Markov approximation. The neglect of these effects for weak coupling  is often referred to as the Born approximation.  If on top of this, the free system evolution (except for an oscillating phase),  is on a time-scale much slower than the memory-time, $\Gamma T_{\rm m}\ll 1$, then  $a(s)$ can be taken to the front of the integral in the damping term of the Langevin equation and the integral over the memory function can be done. The equation for $a(t)$ will then be local in time; this  is referred to as the Markov approximation for the damping.

In our case, the Markov approximation can be made for the damping if the memory-time, $T_{\rm 
m}$,  is  much less than 
the time-scale of both the evolution of the trapped mode, $\Gamma^{-1}$ and the damping, $\gamma^{-1}$. The operator $a(s)$ can then be  replaced by its value at $t$ so that  Eq.\,(\ref{langevineqn}) becomes,
\begin{equation}
        \frac{da(t)}{dt}\simeq-\frac{i}{\hbar}[a(t),H_{T}]-i\Delta\omega a(t)-\gamma'a(t)-\sqrt{\gamma}\xi(t),\label{albert}
 \end{equation}
where
 \begin{eqnarray}
\gamma' & = & \gamma\Re\left\{\int^{t}_{-\infty} dsf_{\rm m}(t-s)\right\},\\
\Delta \omega &= &\gamma\Im\left\{\int^{t}_{-\infty} dsf_{\rm m}(t-s)\right\},
\end{eqnarray}
where $\Delta\omega$ is a frequency shift due to the coupling \cite{louisell}.
Equation (\ref{albert}) is equivalent to the master equation for the reduced density matrix of the system \cite{qn}. 

For the condensate mode the oscillation frequency and the decay rate (determined by the spatial mode function) will vary at a rate $\gamma$ with the number of atoms in the mode.   If the collective excitations of the trapped atoms (caused by the change in population) decay much more rapidly than the loss rate then the chemical potential and the condensate mode function determined from the time-independent GPE will be valid on time scales much shorter than those of the loss.   Steck {\em et al.} \cite{steck} have demonstrated that this procedure is valid  by numerically simulating the evolution of the full coupled GPE for the trapped and untrapped fields. In the present case Eq.\,(\ref{albert}) not only describes the damping of the condensate number, it also  determines the approximate evolution of the total quantum state of the condensate mode (assuming a slow phase diffusion \cite{sols,wright,lewenstein}).  

It is likely that a continuous-wave atom laser will also have some form of pumping. If this pumping is replacing atoms at the same rate as they are being removed then it may be possible for the rate of evolution of the spatial mode function of the condensate atoms to be much slower than the rate of coupling. In this case the above  approximation will become  more accurate. 

In this paper we proceed by first assuming that the mode function and the oscillation frequency are constant and then determine the memory time. This memory time is then compared to the time scale  of the dynamics of the trapped mode of interest. If the memory time is much shorter than the time scale of the mode dynamics the above  assumptions are  valid. If this is not the case then we have a fully non-Markovian decay  with a time-varying system frequency and coupling constant. Obviously it is of interest to determine the regions of validity of the two cases.

\subsection{Green's Functions}

To determine the validity of the Markov approximation we need to
 calculate $f_{\rm m}(\tau)$  in the presence of 
the potential $V^{\rm eff}_{U}({\bf x})$ and for an  interaction region 
$\kappa({\bf x})$ . A convenient 
way to do this is via the Green's function introduced above.
The single particle Green's function for the untrapped particles, Eq.\,(\ref{greensfunction}),
can be written in terms of path integrals (see for example, \cite{pathintegrals}) as,
\begin{equation}
        G({\bf x},t;{\bf x}',t')=\int^{{\bf x},t}_{{\bf x}',t'}d{\bf x}(\tau)
        \exp\left\{\frac{iS[{\bf x}(\tau)]}{\hbar}\right\},
        \label{pathintegral}
\end{equation}
where $\tau=t-t'$ and where $S$ is the action of the particle given by,
 $S[{\bf x}(\tau)]=\int^{t}_{t'}d\tau L({\bf x},\frac{d{\bf 
 x}}{d\tau})$, and $L=\frac{1}{2}m(\frac{d{\bf x}}{d\tau})^{2}-V^{\rm eff}_{U}({\bf 
 x})$ is the Lagrangian for the untrapped particles. We are interested in the case where the Lagrangian is the sum 
 of the Lagrangians in each dimension. In this case the Green's 
 function factorizes into three one-dimensional Greens functions. 
 In general, the path integral is difficult to calculate. However, 
 one can make a semi-classical approximation to the  Green's 
 function. It turns out that this approximation is exact for potentials up to 
 quadratic order in the coordinates. In fact, the semi-classical approximation 
 is justified by approximating a particular potential by a quadratic 
 potential \cite{pathintegrals}. In this work we will only deal with quadratic potentials.

In the following sections (Sec.\ref{rf} and \ref{ramansec}) we will consider interaction regions that 
are independently Gaussian shaped in all three dimensions,
so that $\kappa({\bf x})=\kappa(x)\kappa(y)\kappa(z)$, where each $\kappa(j)$, for $j=\{x,y,z\}$, takes the form
 \begin{equation}
        \kappa(j)=\frac{1}{(\sigma^{2}_{j}\pi)^{\frac{1}{4}}}\exp\left\{-\frac{j^{2}}{2\sigma_{j}^{2}}\right\},
        \label{interactionregion}
 \end{equation}
 where $\sigma_{j}$ is the width of the Gaussian in the $j$th dimension. 
 Note that we are ignoring any multiplicative constant of  $\kappa({\bf x})$ as we are interested in the relative fall off of the memory function. The  assumption of a Gaussian interaction  region allows us to calculate memory functions  exactly in many situations.
We do not expect the exact shape of the interaction region to affect the order-of-magnitude estimates for the memory time that we make in this paper. 

\subsection{Properties of the output-coupled atoms}\label{measurements}
In the previous sections we have made some general considerations
about the Markov approximation for the damping of the trapped
mode. There is, however, another aspect to the problem; that of
determining the properties of the output-coupled atoms.  For example,
let us assume that to a good approximation the system exhibits
Markovian damping and the atoms leave the interaction region with a
reasonably well-defined momentum.  Due to the dispersive nature of the
vacuum for atoms  the properties of the untrapped field will have a
non-trivial dependence on position. Very close to the trap, atoms will
not have traveled very far and dispersive effects may not be large.
However, if the atoms experience a lot of dispersion  then the
properties of the field  will correspond to properties of the trapped
mode averaged over some time. Let us assume that there is a
measurement device localized about a position ${\bf x_{0}}$ that is
making destructive measurements (destructive in the sense that the
detector scatters  atoms into free modes far from those of interest;
examples of such detectors are a hot wire or ionization by a laser) on
any atoms that interact with it. In this case there will be an
uncertainty in the time of emission of an atom that is detected at
${\bf x_{0}}$. The description of a continuous measurement process
becomes much more complicated in this regime \cite{myself}. We call
this the non-Markovian regime for the measurements. 

To describe such a situation (see Appendix A) we can define a response function of the system to a particular measurement device, (in analogy with the memory function), as
\begin{equation}
h_{\chi}(t-t')=\int d{\bf x}d{\bf x}'\:\chi({\bf x-x}_{0})\kappa({\bf x}',t')G({\bf x},t;{\bf x}',t'), \label{resp} 
\end{equation}
where $\chi({\bf x-x}_{0})$ describes the spatial extent of the detector. Ideally, the spatial extent of the measuring device will be smaller than that of the interaction region, otherwise  much of the uncertainty will be introduced by the detector itself. $h_{\chi}(t-t')$ is the probability amplitude for a particle that is emitted in the interaction region at time $t'$ to be detected at time $t$ by the detector.

We can define a memory time for the detection as the time interval between the earliest and the latest time that a particle could have been emitted.
The memory time for measurements of the output can be analyzed in the same way as for the damping. If the memory time corresponding to this response function is much shorter than the time scale of the system dynamics then we can make the Markov approximation for the measurements. If this holds then  a detection time can be considered to correspond exactly to an emission time and all the moments of the measured field are proportional to those of the trapped mode.  This will often be a stronger condition than that for Markovian damping.  

\section{Radio-frequency output coupler}\label{rf}

\begin{figure}
\begin{center}
\epsfig{file=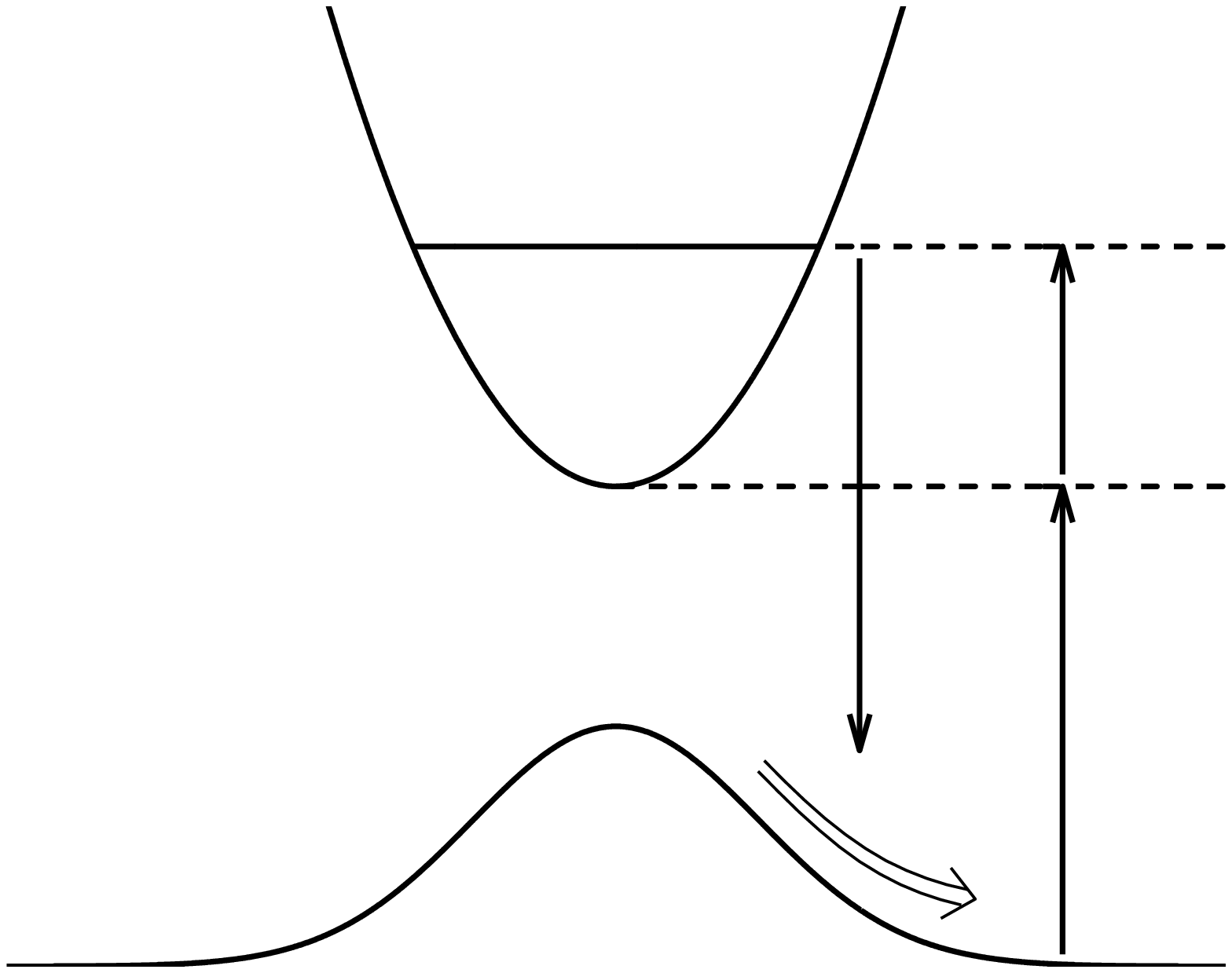,height=0.5\linewidth,width=0.6\linewidth}
\end{center}
\begin{picture}(0,0)

\put(286,120){$\bf\Huge{\omega_{\rm rf}}$}
\put(352,175){$\bf\Huge{\mu}$}
\put(350,80){$\bf\Huge{V_{0}}$}
\put(162,205){$\bf\Huge{ V_{T}({\bf x})}$}
\put(162,60){$\bf\Huge{V^{\rm eff}_{U}({\bf x})}$}
\end{picture}

\caption{\protect \label{schematic} Schematic of the situation under consideration. Atoms are coupled out of a trap by the r.f. field. Once coupled out the atoms see a non-confining potential that tends to repel them from the interaction region.}
\end{figure}

In the radio-frequency output coupler a radio wave of frequency $\omega_{\rm rf}$ induces transitions between  trapped and untrapped (or anti-trapped) magnetic sub-levels of the atoms. The strength of the coupling is given by the Rabi frequency, $\Omega=g\mu_{\rm Bohr}|B|/\sqrt{2}\hbar$, written here in terms of the magnetic field $B$ and the Lande $g$-factor.  The waist of the r.f. wave is assumed to be much broader  than the spatial mode function of the trapped mode and so from Eq.\,(\ref{kappa}) the interaction region becomes $\kappa({\bf x},t)=u_{a}({\bf x})e^{i\nu t}$, where, if $a$ is the condensate mode, $u_{a}({\bf x})$ is determined by the solution to the time-independent GPE. However, we assume here that it is valid for our purposes to approximate this mode function by a Gaussian.  The energy difference between the untrapped level and the center of the trap is given by  $V_{0}=V_{T}(0)$.  The untrapped  atoms are free to propagate away and are in general subject to accelerating potentials, $V^{\rm eff}_{U}({\bf x})$.  The general situation is depicted in Fig.\,\ref{schematic}.

In the following sections we will determine memory times for a number of relevant potentials for the untrapped atoms.

\subsection{Free space}
In order to emphasize the effect of the external potentials we first consider the case when the atoms are coupled into free space. 
In the $x$ dimension the free space, ($V^{\rm eff}_{U}({\bf x})=0$), Green's function is 
\begin{equation}
        G(x,x';\tau)=\frac{1}{\sqrt{4 \pi i  \lambda\tau}}
        \exp\left\{i\frac{(x-x')^{2}}{4\lambda\tau}\right\},
        \label{freespace}
\end{equation}
where $\lambda=\hbar/2m$ is a measure of the rate of spreading of the wave packet. Integrating this over the Gaussian integration region and multiplying the three integrals for each dimension together yields the 
memory function
\begin{equation}
        f_{\rm m}(\tau)=\prod_{j}\Lambda_{j}(\tau)
e^{-i\omega_{0}\tau},
        \label{freespacememoryfun}
\end{equation}
where $j=\{x,y,z\}$, $\omega_{0}=\mu+V_{0}-\omega_{\rm rf}$ and,
\begin{equation}
\Lambda_{j}(\tau)=\frac{\sigma_{j}}{\sqrt{\sigma^{2}_{j}+i\lambda\tau}}.
\end{equation}
A Gaussian wave packet will keep its Gaussian shape (in real space) but will increase in width over time due to the fact that it contains a range of velocity components. The  overlap of the wave-packet with itself as a function of  time is given by the memory function.  

The radio frequency field couples an atom from the trapped mode to modes of the untrapped field with frequencies around $\omega_{0}=V_{0}+\mu-\omega_{\rm rf}$ so that $v=\sqrt{2\hbar\omega_{0}/m}$ is the magnitude of the mean  velocity of the output coupled atoms. When the radio frequency field is on resonance with the trapped mode, $(\omega_{0}= 0)$, particles are coupled out with a zero mean velocity. The particles can only leave the interaction region by quantum mechanical spreading of their wave packets. The memory function will then decay
as $\tau^{-3/2}$ for long times, $\tau\gg \sigma_{j}^{2}/\lambda$. If the output coupled particle has an initial mean velocity, $\omega_{0}\neq 0$, the memory function will decay at a faster rate, as the atom will leave the interaction region more quickly. The case where the mean kinetic energy of the atom is much higher than the coupling rate, $\omega_{0}\gg\gamma$, is very similar to the optical case and a memory time can be defined for which $\gamma^{-1}\gg T_{\rm m}\gg \omega_{0}^{-1}$.

Let us investigate the memory time more  quantitatively.  In order to compare memory functions which decay in very different ways we consider the ratio of the magnitude of the two integrals in Eq.\,(\ref{abs}),
\begin{equation}
R=\frac{\left|{\displaystyle\int^{t-T_{\rm m}}_{-\infty}ds f_{\rm m}(t-s)}\right|}{\left|{\displaystyle\int^{t}_{-\infty}ds f_{\rm m}(t-s)}\right|}.\label{ratio}
\end{equation}
The quantity $R$ is a measure of the inaccuracy of the approximation that a particle has left the interaction region for a particular choice of $T_{\rm m}$. Often we are interested in the inverse, i.e., the memory time given a certain lower bound on the accuracy of the approximation.  Doing the integrals in Eq.\,(\ref{ratio})  we can find the ratio, $R$, in terms of the memory time (see the Appendix B for details of the calculations).
First consider the case when the system is on resonance, $\omega_{0}=0$. In the symmetric interaction region case, $\sigma=\sigma_{j}$, the ratio reduces to
\begin{equation}
R=\frac{\sigma}{[\sigma^{4}+(\lambda T_{\rm m})^{2}]^{1/4}}.\label{la}
\end{equation}
Inverting this equation we find the  memory time in terms of the ratio, $T_{\rm m}\geq \sigma^{2}/\lambda R^{2}$, where we have assumed $R\ll 1$. In the asymmetric case, the memory time is the same as the symmetric case where $\sigma$ is the broadest width of the interaction region.  This memory time depends on the square of the width of the Gaussian and so is a sensitive function of the size of the interaction region. In experiments performed to date the size of a condensate in the broadest dimension has been $\sigma>10\mu$m. For $^{87}$Rb ($m\sim 10^{-25}$kg), if we assume a ratio of $R=10^{-2}$ and a size of $\sigma\sim 10\mu$m, this already gives a very long memory time of the order of $10^{3}$s. 

When the atoms are coupled out with an initial velocity, $\omega_{0}\neq 0$, we get 
\begin{equation}
T_{\rm m}\geq\frac{\sigma^{2}}{\lambda R^{\frac{2}{d}}},\label{pop}
\end{equation}
where $d$ is the dimension of the untrapped field, such that, in the case where the interaction region is cigar shaped, (e.g., $\sigma=\sigma_{x}=\sigma_{y}$ and $\sigma\ll\sigma_{z}$), $d=2$, and  when it is pancake shaped, (e.g. $\sigma_{x}=\sigma_{y}$, and $\sigma_{x}\gg\sigma_{z}=\sigma$), $d=1$.  The initial velocity produces an oscillation of the memory function which when averaged over many oscillations leads to a reduction in the memory time compared to the on resonance case.  For a cigar shaped region with $\sigma\sim 10\mu$m and $R=10^{-2}$ this gives a much shortened (but still relatively long) memory time of $T_{\rm m}\sim 10$s for Rb atoms. These calculations show that atoms of low velocity leaving the interaction region by the spreading of their wave packets linger in the region of interaction for times much longer than $1$s. 

For very weak coupling, the oscillation frequency, $\omega_{0}$, due to the initial kinetic energy of the atoms may be much greater  than the coupling rate $\gamma$. The memory function averaged over times much longer than the time scale defined by the damping $1/\omega_{c}\ll 1/\gamma$ is given by 
\begin{equation}
\overline{f_{\rm m}(\tau)}\propto \frac{\sinh\left(\left[\sigma^{2}+i\lambda\tau\right]\frac{\omega_{c}}{\lambda}\right)}{\sigma^{2}+i\lambda\tau}. 
\end{equation}
An integral over this function converges, as it acts like a ${\rm sinc}(\omega_{c}\tau)$ function for large $\tau\gg\sigma^{2}/\lambda$ and we can define a memory time by $T_{\rm m}\sim 1/\omega_{c}$. Note that we do not analyze the form of the decay in this case as it is simply due to our choice of a sharp cutoff to restrict the frequencies (see Appendix B). 
In this regime, there is essentially no difference between the Markov approximation for the damping in the optical and the atomic case. This is due to the fact that we have implicitly assumed that the coupling constant is approximately constant across the frequencies of interest and that we are on a linear part of the dispersion curve or, equivalently, that the atoms have  a limited range of velocities about a fast mean velocity and so have fast propagation times across the region of interaction. 

\subsection{Gravity}
In most situations atoms coupled out of a trap will be subject to gravitational forces. It is therefore of interest to consider the effects of gravity on the length of the memory time.
  
The Green's function in the $z$-dimension for a gravitational potential $V_{U}^{\rm eff}(z)=-gmz$  has the form
\begin{equation}
                G(z,z';\tau)=\sqrt{\frac{1}{4\pi i\lambda\tau}}
        \exp\left\{i\frac{(z-z')^{2}}{4\lambda\tau}
        -i\frac{gm(z+z')}{2\hbar}\tau-i\frac{mg^{2}\tau^{3}}{24\hbar}\right\}.
\label{linearpotential}
\end{equation}
The first term in the exponential being the usual dispersion term and  the second and third terms can be recognized as phase shifts due to the potential and kinetic  energies, respectively. Integrating over a Gaussian shaped interaction region and assuming free space Green's functions for the other two dimensions the memory function becomes
\begin{equation}
        f_{\rm m}(\tau)=\prod_{j}\Lambda_{j}(\tau)\exp\left\{
        -\left(\frac{mg\sigma_{z}\tau}{2\hbar}\right)^{2}-i\frac{m g^2}{24\hbar}\tau^3  -i\omega_{0}\tau\right\},\label{gravmfun}
\end{equation}
where $\hbar\omega_{0}=\hbar\mu+V_{0}-\hbar\omega_{\rm rf}$ is the initial energy of the output coupled particles (measured from $V_{U}^{\rm eff}(z=0)=0$).
Under the influence of gravity an initial Gaussian wave packet preserves its Gaussian shape (in real space) but the peak of the Gaussian propagates at a velocity $v=gt$ after a time $t$ in the $-z$ direction.
  The Gaussian decay of the memory function (given by the first term in the exponential) is due to the gravitational potential accelerating particles out of the interaction region.  If we assume that this Gaussian decay is the dominant process for short times the memory function can be written as
\begin{equation}
f_{\rm m}(\tau)\propto\exp\left\{-\frac{\tau^{2}}{2\sigma_{\tau}^{2}}\right\},
\end{equation}
where $\sigma_{\tau}=\sqrt{2}\hbar/mg\sigma_{z}$. The ratio, $R$, defined by Eq.\,(\ref{abs}), becomes
\begin{equation}
R\approx\frac{\sigma_{\tau}\sqrt{2}e^{-\frac{T_{\rm m}^{2}}{2\sigma_{\tau}^{2}}}}{\sqrt{\pi}T_{\rm m}},
\end{equation}
where we have used the asymptotic behavior of ${\rm erfc}(x)$ \cite{mathfunction}.
Inverting this, and assuming $T_{\rm m}\gg \sigma_{\tau}$, we find $T_{\rm m}\geq\sigma_{\tau}\sqrt{2\ln(1/R)}=2\hbar\sqrt{\ln(1/R)}/mg\sigma_{z}$. In this case, $T_{\rm m}$ is  inversely proportional to the size of the interaction region in the $z$-direction. Surprisingly, the time for a particle to leave the interaction region actually gets shorter as the interaction region gets larger. For an interaction region of size $\sigma_{z}\sim 10\mu$m this Gaussian envelope gives a memory time of $T_{\rm m}\sim 10^{-5}$s for Rb and $T_{\rm m}\sim 10^{-4}$s for $^{23}$ Na and will become shorter for a larger interaction region. The memory time is largely independent of $R$ for $R\ll 1$, as, in this case, $\sqrt{\ln(1/R)}\sim 1$.

Let us now consider the second term in the exponential in Eq.\,(\ref{gravmfun}). This term describes the property that after a certain time the 
particle is traveling at a velocity high enough that it causes a rapid oscillation of the memory function. This oscillation will average to zero over a time scale much longer than the oscillation frequency. It makes physical sense to estimate  a memory time as the time it takes a particle, accelerating from rest under gravity,  to reach a velocity that is high enough that the particle can no longer be coupled back into the trap, $t=v/g$, where $v\gg\sqrt{2\hbar\gamma/m}$. From Eq.\,(\ref{gravmfun}) the memory function will be oscillating much faster than $\gamma$ if $T_{\rm m}$ is such that 
\begin{equation}
2\pi\frac{mg^{2}T_{\rm m}^{2}}{24\hbar}\gg\gamma.
\end{equation}
Rearranging this, we find $T_{\rm m}\gg 2/g\sqrt{\gamma\hbar/m}$ which supports our initial estimate quite well.  Note that this memory time is independent of the size of the interaction region. For Rb  and assuming the oscillation is 10 times greater than $\gamma$, we can estimate the memory function in terms of $\gamma$ as  $T_{\rm m}\sim 10^{-5}\sqrt{\gamma}$s$^{3/2}$. 

\begin{figure}
\begin{center}
\epsfig{file=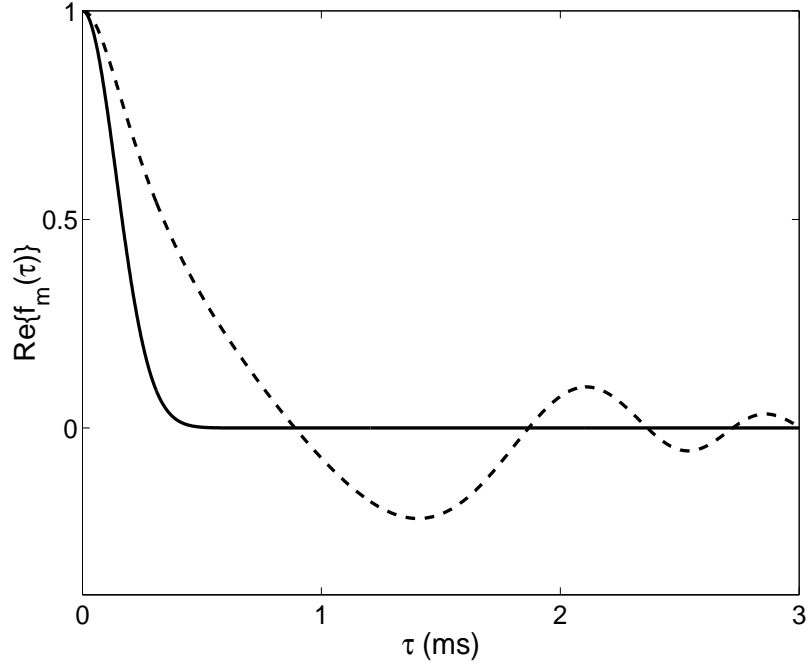,height=0.5\linewidth,width=0.6\linewidth}
\end{center}
\caption{\protect Plot of the real part of the memory function for a particle in a gravitational potential. The memory function is plotted for Sodium atoms with a symmetric interaction region $\sigma_{x}=\sigma_{y}=\sigma_{z}$ and the frequency $\omega_{0}=2\pi\times 100$Hz. The solid line corresponds to $\sigma_{z}=10\mu$m where the time for a particle to leave the spatial interaction region  determines the memory time.  The dashed line depicts the case of a smaller interaction region $\sigma_{z}=1\mu$m where it is possible that  the  oscillation due to the accelerating particles velocity determines the memory time.\label{gravity} } 
\end{figure}

The particle will leave the interaction region sooner than it can gain the required kinetic energy if $\sigma_{\tau}$ is less than the time it takes to make one oscillation $\sigma_{\tau}<(48\pi/mg^{2})^{1/3}$. We can rewrite this condition as $\sigma_{z}>(48 m^{2}g\pi)^{-1/3}$. For Na, $\sigma_{z}$ would need to be greater than  $6\mu$m and for Rb, $\sigma_{z}$ needs to be greater than $0.1\mu$m for the oscillation to become important. These values correspond to very small condensates and therefore we expect the memory time to be determined by the time it takes the particles to leave the interaction region in most situations. A plot of two possible situations is given in Fig.\,\ref{gravity}.

\subsection{Collisional repulsion and the anti-trapped case}\label{collisions}
If particles are being coupled out of the condensate mode into an untrapped state they will 
see a repulsive potential due to the condensate atoms left in the trap (in the repulsive interaction case) and the shape of 
this potential will be proportional to the density distribution of the 
condensate. In the Thomas-Fermi approximation the condensate density 
takes the shape of the trap potential \cite{barnett}, $U N|u_{a}({\bf x})|^{2}={\rm 
max}\left[\mu+V_{0}-V_{T}({\bf x}),0\right]$. In the present experimental 
situations the traps have been harmonic in 
all three dimensions. The repulsive potential for the untrapped atoms will then be a three-dimensional inverted harmonic potential, given by
 $V^{\rm eff}_{U}({\bf x})=\epsilon UN|u_{a}({\bf x})|^{2}={\rm max}[\epsilon\mu-\frac{1}{2}m\sum\epsilon(\omega^{T}_{j})^{2} j^2,0]$, where $\epsilon$ is the ratio of the scattering lengths between a trapped-untrapped atomic collision and a trapped-trapped collision. In this paper we assume that in the region of interaction we can approximate the effect of the cutoff inverted harmonic potential by a inverted harmonic potential that is not cutoff and is therefore quadratic everywhere.  In general, the Green's function depends on the potential everywhere, but we are only interested in the time until a particle is repelled from the interaction region and as long as the particle energies are not too close to the cutoff we can neglect the global effects due to the shape of the potential outside the interaction region. 

In addition, if we are considering particles that are output 
coupled into spin states that are repelled by the trap this is also 
an inverted harmonic potential. In this case the repulsive potential will be $V^{\rm eff}_{U}({\bf x})=\epsilon UN|u_{a}({\bf x})|^{2}-V_{T}({\bf x})$, where we have also included the collisional repulsion. In this case, $\epsilon$ is the ratio between a trapped-antitrapped atomic collision and a trapped-trapped collision.

 The Green's function for 
an inverted harmonic potential is easily determined from the Green's function for 
an harmonic potential \cite{pathintegrals} with the substitution $\omega\rightarrow i\omega$. 
Along a single axis of the inverted harmonic potential we have
\begin{equation}
                G(x,x';\tau)=\sqrt{\frac{m\omega_{x}}{2\pi i\hbar\sinh\omega_{x}\tau}}
\exp\left\{\frac{im\omega_{x}}{2\hbar\sinh\omega_{x}\tau}\left[(x^{2}+x'^{2})
        \cosh\omega_{x}\tau-2xx'\right]-i\frac{\tilde{V_{0}}}{\hbar}\tau\right\},
\label{inverted harmonicpotential}
\end{equation}
where $\tilde{V_{0}}$ is the potential at the center of the inverted harmonic. This is the Green's function of the untrapped atoms if we can assume that the effect of the inverted harmonic potential is much greater than that of gravity. In light of the above discussion it is rather inconsistent now to assume a Gaussian interaction region. However, we do not expect the exact shape of the interaction region to effect our results dramatically. This is borne out by results which we will present below.

The corresponding memory function in the Gaussian case becomes
\begin{equation}
        f_{\rm m}(\tau)=\prod_{j}\left[\left(\cosh\omega_{j}\tau+
        i\left(\frac{\lambda}{\omega_{j}\sigma_{j}^{2}}-\frac{\omega_{j}\sigma_{j}^{2}}{\lambda}\right)\sinh\omega_{j}\tau\right)\right]^{-1/2}e^{-i\omega_{0}\tau},
\end{equation}
where  $\omega_{0}=\mu+V_{0}/\hbar-\omega_{\rm rf}-\tilde{V_{0}}/\hbar$.
For times $\tau\gg 1/\omega_{j}$ this becomes
\begin{equation}
        f_{\rm m}(\tau)\simeq 
\prod_{j}\left[\left(1+i\frac{\lambda}{\omega_{j}\sigma_{j}^{2}}-i\frac{\omega_{j}\sigma_{j}^{2}}{\lambda}\right)\right]^{-1/2}\exp\left\{-\frac{\omega_{j}}{2}\tau-i\omega_{0}\tau\right\}.\label{short}
\end{equation}
This exponential decay describes particles being repelled out of the interaction region by the inverted harmonic potential. Assuming this exponential decay is the dominant process for short times we can estimate the usual ratio of integrals by  $R\approx\exp(-3\bar{\omega}T_{\rm m}/2)$, where $\bar{\omega}=(\omega_{x}+\omega_{y}+\omega_{z})/3$ is the  mean of the inverted harmonic trap frequencies. Inverting this we get $T_{\rm m}\geq(2/3\bar{\omega})\ln(1/R)$, which is independent of the size of the interaction region.  For very asymmetric traps the
trap frequency with the largest magnitude will define the memory time as particles will be repelled fastest in this dimension.

In the above calculations we assumed a Gaussian interaction region. However, these results hold for general interaction regions. This can be seen by taking the limit $\tau\gg 1/\omega_{j}$ (as suggested by Eq.\,(\ref{short})) of the Green's function itself,
\begin{equation}
    G(x,x';\tau\gg 1/\omega_{j})=\sqrt{\frac{m\omega_{x}}{2\pi i\hbar}}
\exp\left\{\frac{im\omega_{x}}{2\hbar}(x^{2}+x'^{2})-\frac{\omega_{x}}{2}\tau-i\frac{\tilde{V_{0}}}{\hbar}\tau\right\},
\end{equation}
where we have assumed $1\gg e^{-\omega_{x}\tau}$. This decays as an exponential with increasing time, independent of the shape of the interaction region.

In the untrapped case where the potential is determined by collisional repulsion  $\omega_{j}=\sqrt{\epsilon}\omega_{j}^{T}$ and $\tilde{V_{0}}=\epsilon\mu$. Assuming that $\epsilon\simeq 1$  we can estimate the memory time (as defined by the exponential decay) as simply one over the  mean of the trap frequencies. Typical trap frequencies range from $2\pi\times 10-2\pi\times 400$ Hz. From the experimental parameters of \cite{mewes} we can determine the value  $\bar{\omega}=2\pi\times 219$Hz for the average trap frequency which  gives a memory time of  $T_{\rm m}\sim 1$ms. 

In the anti-trapped case we have $\omega_{j}=(\sqrt{\epsilon}+1)\omega^{T}_{j}$ and $\tilde{V_{0}}=\epsilon\mu -V_{0}$. The effective inverted harmonic trap potential is estimated to be approximately twice the strength of the trap potential as it is the sum of the  repulsive  potential and the collisional repulsion. So the memory time calculated for the untrapped case will be halved due to the repulsive trap potential.

\subsection{Summary of the radio-frequency output coupler}\label{rfsummary}

A summary of the calculated memory times for the radio-frequency output coupler is given in Table\ \ref{table}, where $R\gg 1$ and $d$ is the dimension of the space of the untrapped atoms, e.g., $d=1$ corresponds to a cigar shaped interaction region.

If particles are coupled out with a large kinetic energy, $\omega_{0}\gg\gamma$, then a memory time can be defined by $\gamma^{-1}\gg T_{\rm m}\gg \omega_{0}^{-1}$, in the same way as for an optical system. In the optical case the field inside a cavity is coupled directly to the field outside and particles tunneling out of the cavity have nearly the same energy as they had inside  the cavity. However, in the atomic case the coupling between the trapped field and the untrapped field is mediated by the radio-frequency field. Particles that make the transition between a trapped state and an untrapped state have an initial energy equal to the detuning between the frequency of the r.f. field and the energy difference between the two states, see Fig.\,\ref{schematic}.  This means that in the case where $\omega_{0}$ is large the radio-frequency is far from resonance  with the condensate mode. If a populated non-condensate mode is close to resonance  the particles in the output will be thermal particles. In order to maximize the coupling to the condensate, the r.f. field will need to be on resonance with the condensate mode and consequently we expect particles to come out with a small mean energy, $\omega_{0}\approx 0$. The slow moving particles are then subject to the accelerating potentials of gravity and collisional repulsion. 

From our calculations, gravity seems to be the force that determines the memory time. It yields a memory time of $T_{\rm m}\sim 10^{-2}$ms for a Rb condensate with a width in the $z$-direction of the order of  $10\mu$m and decreasing with increasing width. Sodium, being lighter, has a longer memory time. Whereas collisional repulsion gives a memory time of the order of the inverse trap frequencies  $T_{\rm m}\sim 10^{-1}$ms. This depends on the scattering length of a trapped-untrapped collision and is independent of the size of the region of interaction.

\section{Raman output coupler} \label{ramansec}
We can model the case of a Raman output-coupler by replacing the 
coupling constant in the direction of the kick, $\kappa(x)$, by 
 \begin{equation}
        \kappa(x)\rightarrow\kappa(x)e^{-i\Delta x},
        \label{ramankick}
 \end{equation}
where $\hbar\Delta$ is given by the difference between the momentum of the 
two photons involved in the Raman transition, $\Delta=k_{1}-k_{2}$.

Assuming the Raman kick is the dominant process, the Green's function 
is the free space Green's function. The memory function,
assuming a Gaussian shaped interaction region, is then
\begin{equation}
        f_{\rm m}(\tau)=\prod_{j}\Lambda_{j}(\tau)
        \exp\left\{-\frac{i}{2}\frac{\lambda\sigma^{2}_{x}\Delta^{2}\tau}{\sigma^{2}_{x}+i\lambda\tau}-i\omega_{0}\tau\right\},
        \label{ramanmemfun}
\end{equation}
where  in this case $\omega_{0}=\mu-\Delta\omega+V_{0}/\hbar$, and we have defined $\Delta\omega=\omega_{1}-\omega_{2}$ as the difference between the frequencies of the two lasers.

\begin{figure}
\begin{center}
\epsfig{file=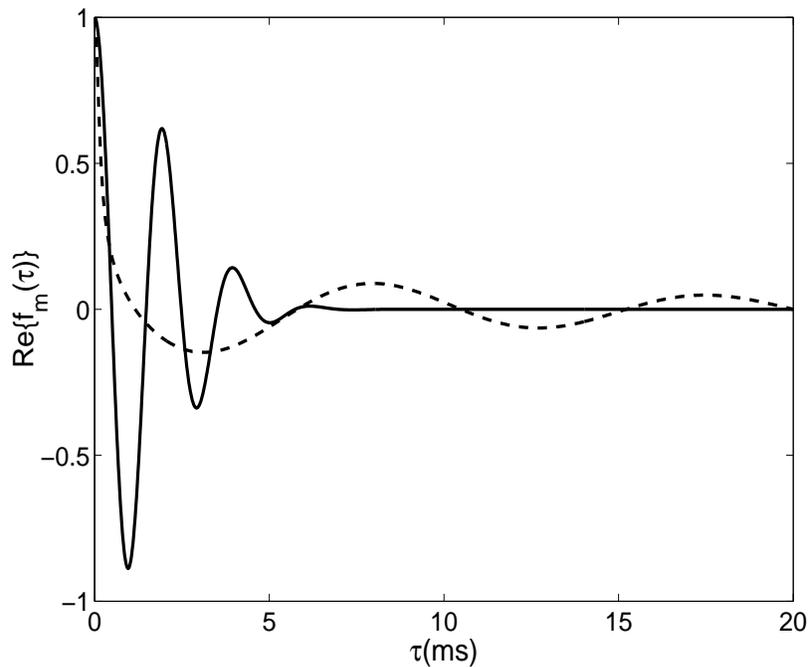,height=0.5\linewidth,width=0.6\linewidth}
\end{center}
\caption{\protect\label{raman}  This figure is a comparison between the case when a particle is given a well defined momentum kick and when it is not for the Raman output coupler. We have plotted the real part of the memory function as a function of time for the parameters $\Delta=10^{6}$m$^{-1}$, $\sigma_{y}=\sigma_{z}=10\mu$m and $\omega_{0}=2\pi\times 100$Hz. The solid line  is the case $\sigma_{x}=10\mu$m so that $\Delta\gg 1/\sigma_{x}$. The dotted line is the case $\sigma_{x}=1\mu$m, so $\Delta\sim 1/\sigma_{x}$. In this last case the memory function does not exhibit the Gaussian decay as the momentum kick is not well defined. }
\end{figure}

 In this case it is possible for focused laser beams to be of the order of the size of the condensate. The width of these beams would then  define the size of the interaction region. However, focusing the beams to a small region will tend to lead to diffraction. The momentum of the kick will not be well defined in this case. A comparison  between the situation when the momentum kick is well defined and when it is not is shown in  Fig.\,\ref{raman}. If the momentum kick is not very large compared to the characteristic inverse length of the interaction region, $\Delta\sim 1/\sigma_{x}$, then for long times the behavior is similar to the radio frequency case (i.e., repulsive potentials will dominate). However, where the kick is much greater than the inverse size of the region $\Delta \gg 1/\sigma_{x}$ we can expand the term in the exponential in Eq.\,(\ref{ramanmemfun}) and write the memory function as
\begin{equation}
        f_{\rm m}(\tau)\approx\prod_{j}\Lambda_{j}(0)
        \exp\left\{-\frac{\tau^{2}}{2\sigma^{2}_{\tau}}-i(\lambda\Delta^{2}+\omega_{0})\tau\right\},\label{goonies}
 \end{equation}
where $\sigma_{\tau}=\sigma_{x}/\sqrt{2}\Delta\lambda$. This will dominate for  
time scales $\tau\ll \sigma_{x}^{2}/\lambda$.   In the case when the kick is well defined the memory time has a very simple classical interpretation in terms of the time a particle takes, $t=d/v$, to cross a 
distance $d=2\sigma$ going at velocity $v=\hbar\Delta/m$.
The first term in Eq.\,(\ref{goonies}) is a Gaussian envelope of width $\sigma_{\tau}$ and represents the overlap between the wave packet of a  particle propagating away and the interaction region.   As in the gravitational case this Gaussian envelope determines a memory time in terms of the usual ratio of integrals as
$T_{\rm m}\geq \sigma_{x}\sqrt{\ln(1/R)}/\Delta\lambda$ .  In the best-case scenario, if the two laser beams involved in the Raman transition are counter-propagating, then at optical frequencies the kick given to the atom will be of order $\Delta\sim 10^{7}$m$^{-1}$. This  yields a memory time of $T_{\rm m}\sim 10^{-3}$s for interaction regions of $\sigma_{x}=10\mu$m for Rb atoms.  

The second term in Eq.\,(\ref{goonies}) is recognized as a kinetic energy term for a particle 
propagating with momentum $\hbar(\Delta^{2}/\lambda+\omega_{0})$. If the kick is large the particle will be given a large kinetic energy and a memory time can be determined by a high oscillation frequency. If $\Delta^{2}\lambda+\omega_{0}\gg\gamma$, then  we can determine a memory time as $T_{\rm m}\sim 1/\omega_{c}$, where $\Delta^{2}\lambda+\omega_{0}\gg\omega_{c}\gg \gamma$. Unlike the radio frequency output coupler the Raman output coupler (in the case where the kick is well defined) is very similar to the laser in that condensate atoms will  leave the trap with a finite, and possibly large, kinetic energy in a well-defined direction. If the kick is large enough the effects of gravity may be negligible for short distances.
 
\section*{Discussion}

We have investigated the regimes of validity of the Markov approximation for  atoms that are being coupled out of an atomic trap by an  output coupler by  determining  the memory times (or correlation times) of the output coupled atoms. A memory time can be defined by the time taken for a  particle to leave the region of phase space of the untrapped field where it can be coupled back into the trap. After this time, if a particle has not been coupled back it can be considered to have irretrievably left the trap.

Atoms coupled out by an r.f. field that leave  the interaction region by the free space spreading of their atomic wave packet alone can have long memory times, ($T_{\rm m}\gg 1$s), which are dependent on the square of the size of the broadest dimension of the interaction region. These memory times are reduced if atoms are coupled out with a large mean kinetic energy. However, in this case the coupling will be far from resonance with the condensate mode. 
For the r.f. output coupler, gravity (which will nearly always be present) dominates in many cases, and yields a memory time that ranges from $ 10^{-2}-10^{-1}$ms for typical experimental parameters.  Collisional repulsion  leads to a memory time of the order of the inverse mean trap frequency, $\bar{\omega}$, and depending on the ratio of the scattering lengths between a trapped-untrapped and a trapped-trapped collision, $\epsilon$: $T_{\rm m}\sim 1/\sqrt{\epsilon}\bar{\omega}$, which is slightly longer than the memory time for gravity for typical experimental parameters.

Raman output coupling has a memory time which depends on the inverse strength of the momentum kick, $\hbar\Delta$, given by the light to the atoms and on the size of the condensate in the direction of the kick, $\sigma$. The kick must be well-defined, $\Delta\gg 1/\sigma$, to produce a reduction of the memory time.  If the light beams are focused too tightly ($\Delta\sim 1/\sigma$) diffraction effects will dominate and the Raman output coupler, like the radio-frequency output coupler, must rely on external potentials such as gravity to determine a memory time.  In the best case scenario, two counter propagating laser beams will produce a kick of $\Delta\sim 10^{7}$m$^{-1}$ giving a memory time of $T_{\rm m}\sim 1$ms. The Raman output coupler has the nice property that in the regime where a short memory time is produced the atoms are given a well-defined momentum kick, producing a beam of atoms \cite{raman}.
 
These memory times must be short compared with the time-scales of the system evolution, $\Gamma^{-1}$, and the coupling rate $\gamma^{-1}$, in order to make the Markov approximation. The most important upper time limit on $\gamma^{-1}$ is the correlation time of the condensate, $\tau_{c}$. $\gamma^{-1}$ must be short compared to $\tau_{c}$ otherwise the output-coupled atoms will not be correlated with each other. Current experimental estimates for the correlation time give a range of from $0.1-1$s, i.e., long compared to the calculated memory time corresponding to gravity. Therefore, the Markov approximation will be valid for a range of coupling rates where $\tau_{c}\gg \gamma^{-1}\gg T_{\rm m}$.   

 When the Markov approximation is valid a Markov master equation may be used to solve for the evolution of the trap modes undergoing damping.  When the Markov approximation cannot be made new methods must be employed \cite{hope97,myself,gisin}, and novel behavior will be observed \cite{moy,savage,hope98}.  

\vspace{1cm}

This work was carried out with the help of funding from the Marsden Fund of the Royal Society of New Zealand, the University of Auckland Graduate Research Fund and the USA/NZ Cooperative Research Program. M.N. acknowledges financial support by the Deutsche Forschungsgemeinschaft.
M.J. would like to thank Professor Roy Glauber for his hospitality  during M.J's stay at Harvard University where this work was conceived, and J. Ruostekoski and J. Longdell for helpful discussions.
\section*{Appendix A}
A localized system (in this case the trapped atoms) interacting with a bath (the modes of the untrapped field) can be thought of in terms of inputs and outputs to the system \cite{collett}. The initial bath  propagates towards the system; interacts with the system; propagates away again; and is eventually measured. To formalize this idea, input and output fields are defined by 
\begin{equation}
   \hat{\psi}_{\rm in}({\bf x},t)=U^{\dagger}_{U}(t,t_{0})\hat{\psi}_{U}({\bf x},t_{0})U_{U}(t,t_{0}),
\end{equation}
where $t_{0}<t$ and is usually taken to be in the distant past, and
\begin{equation}
   \hat{\psi}_{\rm out}({\bf x},t)=U_{U}(t_{1},t)\hat{\psi}_{U}({\bf x},t_{1})U^{\dagger}_{U}(t_{1},t),
\end{equation}
where $t_{1}>t$ and is usually taken to be in the distant future, and where the field operators are in the Heisenberg picture. $U_{U}$ is the evolution operator defined in terms of $H_{U}$ alone.
 The relation between the input and output fields and the trap mode is given by
\begin{equation}
 \hat{\psi}_{\rm out}({\bf x},t)=  \hat{\psi}_{\rm in}({\bf x},t)+\sqrt{\gamma}\int^{t}_{t_{0}}ds [\hat{\psi}_{\rm in}(t,{\bf x}),\hat{\xi}^{\dagger}(s)] a(s).
\end{equation}
The output has a contribution from the input field and the trapped mode at earlier times.

We assume that our measuring device (of Sec.\ref{measurements}) measures normally ordered moments \cite{glauber} of the quantity
\begin{equation}
\hat{\Psi}_{\rm out}(t) =\int d{\bf x} \chi({\bf x-x}_{0})\hat{\psi}_{\rm out}({\bf x},t),
\end{equation}
where $\chi({\bf x-x}_{0})$ describes the spatial extent of the detector. We are assuming that the actual measurements take place over a very short time. The response function for the system given a particular measurement device is defined as
\begin{equation}
h_{\chi}(t-t')=\int d{\bf x}\chi({\bf x-x}_{0}) [\hat{\psi}_{\rm in}(t,{\bf x}),\hat{\xi}^{\dagger}(t')].
\end{equation}
$h_{\chi}(t-t')$ is the probability amplitude for a particle that is emitted in the interaction region at time $t'$ to be detected at time $t$ by the detector. This becomes more obvious if we write it in terms of the Green's function for the untrapped field Eq.\,(\ref{resp}).

We can define a memory time, $\tilde{T}_{\rm m}$, as the time interval between the earliest time a detected particle could have been emitted and the latest time the particle could have been emitted. $\tilde{T}_{\rm m}$ exists if we can write
\begin{equation}
\int_{t_{0}}^{t} ds h(t-s) a(s)\simeq \int^{t}_{t-\tilde{T}_{\rm m}} ds h_{\chi}(t-s)a(s),
\end{equation}
where we have ignored  any constant time delay that the response introduces  as it will not effect steady-state results. If this memory time is much shorter than the time scales of the system dynamics we can write
\begin{equation}
 \hat{\Psi}_{\rm out}(t)\simeq\hat{\Psi}_{\rm in}(t)+\sqrt{\tilde{\gamma}}a(t), 
\end{equation}
where $\tilde{\gamma}$ is defined in a similar way to $\gamma'$ for the damping and $\Psi_{\rm in}(t)$ is the contribution of the input field to the output. If this holds  a detection time corresponds exactly to an emission time and all the moments of the measured field are proportional to those of the trapped mode.

\section*{Appendix B}

The ratio of the magnitude of the integrals over the memory function from $t-T_{\rm m}$ to $-\infty$ and from $t$ to $-\infty$ in the free space case is
\begin{equation}
R=\frac{\left|{\displaystyle\int^{t-T_{\rm m}}_{-\infty}ds \prod_{j}\Lambda_{j} (t-s)}\right|}{\left|{\displaystyle\int^{t}_{-\infty}ds \prod_{j}\Lambda_{j}(t-s)
e^{-i\omega_{0}(t-s)}}\right|}.\label{ratio2}
\end{equation}
Doing the integrals in this equation allow us to determine the ratio in terms of the memory function.

In the symmetric interaction region case, where $\sigma=\sigma_{j}$, we can do the integrals in Eq.\,(\ref{ratio2})  and we find 
\begin{equation}
R=\left|\frac{{\displaystyle\frac{\sigma e^{-i\omega_{0}T_{\rm m}}}{\sqrt{\sigma^2+i\lambda T_{\rm m}}}-\sqrt{\frac{\omega_{0}\sigma^{2}\pi}{\lambda}}e^{\frac{\sigma^{2}\omega_{0}}{\lambda}}}{\rm erfc}\left\{\sqrt{\omega_{0}\left(\frac{\sigma^{2}}{\lambda}+iT_{\rm m}\right)}\right\} }{{\displaystyle 1-\sqrt{\frac{\omega_{0}\sigma^{2}\pi}{\lambda}}e^{\frac{\sigma^{2}\omega_{0}}{\lambda}}}{\rm erfc}\left\{\sqrt{\frac{\omega_{0}\sigma^{2}}{\lambda}}\right\}} \right|,\label{symratio}
\end{equation}
where ${\rm erfc}(z)=1-{\rm erf}(z)$ is the complementary error function. In the case where $\omega_{0}=0$  this ratio reduces to Eq.\,(\ref{la}). In the asymmetric case with $\omega_{0}=0$,  although the intermediate behavior is governed by the narrowest dimensions, the memory time is determined by the broadest dimension. This is due to the fact that in strictly one and two dimensions we cannot define a memory time as in these cases the integrals in Eq.\,(\ref{ratio}) diverge. 
 This  is a fundamental property that arises from the $1/\sqrt{\tau}$ dependence of the one-dimensional Green's function.  In conclusion, the third dimension is required for the system to be dissipative in the $\omega_{0}= 0$ case. 

On the other hand, the case $\omega_{0}\neq 0$ is important in the asymmetric case as it is well known that a multiplicative oscillating factor can make an otherwise divergent integral convergent. In the  case when $\omega\neq 0$ we can use the asymptotic expansion for the error function, \cite{mathfunction}, $\sqrt{\pi}ze^{z^{2}}{\rm erfc}z\sim 1+1/2z^{2}$ as $z\rightarrow \infty$ for $|{\rm arg}(z)|<3\pi/4$, to show that for long times $T_{\rm m}\gg 1/\omega_{0}$, $R\propto 1/[\sigma^{4}+(\lambda T_{\rm m})^{2}]^{3/4}$. For  $\omega_{0}\gg\lambda/\sigma^{2}$, Eq.\,(\ref{symratio}) simplifies to 
\begin{equation}
R\approx\frac{\sigma^{3}}{\left[\sigma^{4}+(\lambda T_{\rm m})^{2}\right]^{3/4}}.
\end{equation}

In the case where the interaction region is cigar shaped, e.g.,  $\sigma=\sigma_{x}=\sigma_{y}$ and $\sigma\ll\sigma_{z}$,
 the ratio becomes
\begin{equation}
R=\left|\frac{E_{1}\left\{\omega_{0}\left(\frac{\sigma^{2}}{\lambda}+iT_{\rm m}\right)\right\}}{E_{1}\left\{\omega_{0}\frac{\sigma^{2}}{\lambda}\right\}}\right|,
\end{equation}
where $ E_{1}(z)$ is the $1$st order exponential integral \cite{mathfunction}. Note that a one-dimensional interaction region corresponds to a two dimensional bath and vice versa. For long times we can write
$R\propto \sigma^{2}/\sqrt{\sigma^{4}+(\lambda T_{\rm m})^{2}}$, where we have used the asymptotic expansion $E_{1}(z)\approx e^{-z}/z$ as $z\rightarrow \infty$ for $|{\rm arg} z|<3\pi/2$. The ratio simplifies in the $\omega_{0}\gg\lambda/\sigma^{2}$ limit to,
\begin{equation}
R\approx\frac{\sigma^{2}}{[\sigma^{4}+(\lambda T_{\rm m})^{2}]^{1/2}},
\end{equation}
for all times, $T_{\rm m}$. The memory function is then given by $T_{\rm m}\geq \sigma^{2}/R\lambda$.

If the interaction region is pancake shaped, e.g., $\sigma=\sigma_{x}$, $\sigma_{y}\gg\sigma$, and  $\sigma_{z}\gg\sigma$ we get 
\begin{equation}
R=\left|\frac{{\rm erfc}\left\{\sqrt{\omega_{0}\left(\frac{\sigma^{2}}{\lambda}+iT_{\rm m}\right)}\right\}}{{\rm erfc}\left\{\sqrt{\omega_{0}\frac{\sigma^{2}}{\lambda}}\right\}}\right|.
\end{equation}
When $\omega_{0}\gg\lambda/\sigma^{2}$ this becomes
\begin{equation}
R\approx\frac{\sigma}{[\sigma^{4}+(\lambda T_{\rm m})^{2}]^{1/4}}.
\end{equation}
Yielding a memory time with the same dependency as the low frequency symmetric case.
Putting this all together yields Eq.\,(\ref{pop}).

If the atom is coupled out with a large kinetic energy it is useful to consider the memory-function written in the form of an integral over frequency,
\begin{equation}
f_{\rm m}(t-t')=\int^{\infty}_{0} d\omega  D(\omega)|\kappa(\omega)|^{2}e^{-i[\omega-\nu](t-t')},\label{freq} 
\end{equation}
where $\omega=\omega_{\bf k}$, $D(\omega)$ is the density of states and $\kappa(\omega)$ is found by transforming the effective coupling constant to ${\bf k}$-space (where ${\bf k}$ is the label of the modes of the untrapped field) and then using the dispersion relation to make a change of variables to frequency space. 

In dealing with averages over oscillatory functions it is necessary to specify a time scale over which averages are to be taken.  In frequency space this corresponds to considering only a range of frequencies in the integral in Eq.\,(\ref{freq}). A physical frequency-dependent coupling, $\kappa(\omega)$, will naturally limit the range of frequencies and we can consider memory times defined by Eq.\,(\ref{abs}). 

The time scale that we are ultimately interested in though is $\gamma^{-1}$. We should therefore average over time scales short compared to $\gamma^{-1}$.  In frequency space, we can introduce a simple cutoff frequency $\omega_{c}\gg\gamma$ so that the function is averaged over times of order $\omega_{c}^{-1}$,
\begin{equation}
\overline{f_{\rm m}(t-t')}= e^{-i\mu t}\int^{\omega_{c}}_{-\omega_{c}} d\omega D(\omega+\omega_{0})|\kappa(\omega+\omega_{0})|^{2}e^{-i\omega(t-t')}, \label{filteredfun}
\end{equation}
where $\omega_{0}=\mu+\nu >\omega_{c}$.   Obviously, $\omega_{0}$ must be much greater than this for the function to be attenuated with increasing time. We are essentially band filtering the memory function so we refer to this as the filtered memory function. Physically, we are neglecting particles with energy greater than $\hbar(\omega_{0}+\omega_{c})$ and less than $\hbar(\omega_{0}-\omega_{c})$ because they are a long way from resonance.

 We can define  a memory time   in terms of the filtered memory function $\overline{f_{\rm m}(t-t')}$ by
\begin{equation}
\left|\int^{t}_{t-T_{\rm m}} ds\overline{f_{\rm m}(t-s)}\right|  \gg 
\left|\int^{t-T_{\rm m}}_{t_{0}} ds\overline{f_{\rm m}(t-s)}\right|. 
\label{osc}
\end{equation} 
In many cases $T_{\rm m}\sim 1/\omega_{c}$.  Note that a memory time defined in this way relies on the fact that we have assumed $\omega_{0}\gg\gamma$.

The memory function averaged over times $1/\omega_{c}$ where $\omega_{0} \gg \omega_{c} \gg \gamma$ is given by Eq.\,(\ref{filteredfun}). In the symmetric case, $D(\omega)\propto \sqrt{\omega}$ and for a Gaussian shaped interaction region, $|\kappa(\omega)|^{2}=\exp(-\frac{\sigma^{2}}{\lambda}\omega)$. The filtered memory function simplifies for long times to 
\begin{equation}
\overline{f_{\rm m}(\tau)}\propto\sinh\left(\left[\sigma^{2}+i\lambda\tau\right]\frac{\omega_{c}}{\lambda}\right)\left[\frac{\omega_{0}}{\sigma^{2}+i\lambda \tau}+\frac{1}{2(\sigma^{2}+i\lambda\tau)^{2}}\right] , 
\end{equation}
where we have again used the asymptotic expansion for the error function.
The second term will tend more rapidly to zero and so the first term will define the memory time. An integral over this term converges as it acts like a ${\rm sinc}(\omega_{c}\tau)$ function for large $\tau\gg\sigma^{2}/\lambda$ and we can define a memory time by $T_{\rm m}\sim 1/\omega_{c}$. Note that we do not analyze the form of the decay in this case as it is simply due to our choice of a sharp cutoff to restrict the frequencies.

If the interaction region is effectively two dimensional then the density of states becomes $D(\omega)\propto 1/\sqrt{\omega}$  with $|\kappa(\omega)|^{2}$ the same as above. The filtered memory function for long times also simplifies to
\begin{equation}
\overline{f_{\rm m}(\tau)}\propto \frac{\sinh\left(\left[\sigma^{2}+i\lambda\tau\right]\frac{\omega_{c}}{\lambda}\right)}{\sigma^{2}+i\lambda\tau}, 
\end{equation}
where we have again used the asymptotic properties of the error function.
In a cigar shaped interaction region the density of states is $D(\omega)\propto 1$ and the filtered memory function has the same long term behavior as the two and three dimensional cases.

\newpage

\newpage
\begin{table}
\begin{tabular}{l cc}
\rule[.1in]{0cm}{.3cm}
Free space (no oscillation) & ${\displaystyle T_{\rm m}\geq\frac{2m\sigma^{2}}{\hbar R^{2}}}$, & $\sigma={\rm max}(\sigma_{x},\sigma_{y},\sigma_{z})$ \\[.2cm] 
\hline
\rule[.1in]{0cm}{.3cm}
Free space (fast oscillation) & $ {\displaystyle T_{\rm m}\geq\frac{2m\sigma^{2}}{\hbar R^{\frac{2}{d}}}}$, & ${\displaystyle\omega_{0}\gg\frac{\lambda}{\sigma^{2}}}$\\[.3cm]\hline
\rule[.1in]{0cm}{.3cm}
Gravity (Gaussian envelope)&\hspace{.2cm} ${\displaystyle T_{\rm m}\geq \frac{2\hbar}{mg\sigma_{z}}}\sqrt{\ln\left(\frac{1}{R}\right)}$ & \\[.3cm]
 \hline\rule[.1in]{0cm}{.4cm}
Gravity (oscillation) & ${\displaystyle T_{\rm m}\gg \frac{2}{g}\sqrt{\frac{\gamma\hbar}{m}}}$ & \\[.3cm]
 \hline\rule[.1in]{0cm}{.3cm}
Inverted Harmonic (exponential decay)&${\displaystyle T_{\rm m}\geq\frac{6}{\bar{\omega}}}\ln\left(\frac{1}{R}\right)$& \\[.2cm] 
\end{tabular}

\caption{\protect Summary of memory times for the radio-frequency output coupler.\label{table}}
\end{table}
\end{document}